\documentclass[aps,pra,nofootinbib,twocolumn,floatfix,showpacs,superscriptaddress]{revtex4-1}
\usepackage{amsmath,graphicx,epsfig,amsthm,color,bm}
\usepackage{pifont,bbding,amssymb,soul,mathrsfs}

\usepackage{threeparttable}
\usepackage[hyperindex,breaklinks,colorlinks=true,citecolor=blue,urlcolor=blue]{hyperref}
\usepackage{appendix, comment}

\begin{document}

\title{Efficient generation of broadband photon pairs in shallow-etched lithium niobate nanowaveguide }

\author{Xiao-Xu Fang}
\affiliation{School of Physics, State Key Laboratory of Crystal Materials, Shandong University, Jinan 250100, China.}
\affiliation{Shenzhen Research Institute of Shandong University, Shenzhen 518057, China}

\author{Leiran Wang}
\affiliation{State Key Laboratory of Transient Optics and Photonics, Xi'an Institute of Optics and Precision Mechanics, Chinese Academy of Sciences, Xi'an 710119, China}
\affiliation{University of Chinese Academy of Sciences, Beijing 100049,China}

\author{He Lu}
\email{luhe@sdu.edu.cn}
\affiliation{School of Physics, State Key Laboratory of Crystal Materials, Shandong University, Jinan 250100, China.}
\affiliation{Shenzhen Research Institute of Shandong University, Shenzhen 518057, China}


\begin{abstract} 
We design and fabricate shallow-etched periodically poled lithium niobate waveguide to realize highly-efficient broadband spontaneous parametric down-conversion~(SPDC) on nanophotonic chip. The shallow-etched waveguide is capable to tolerate the non-uniformities of waveguide width induced by fabrication imperfections, enabling generation of photon pairs with high count rate and bandwidth. We demonstrate photon-pair generation with a high brightness of 11.7~GHz/mW and bandwidth of 22~THz, in a 5.7-mm-long PPLN waveguide. The generated photon pairs exhibit strong temporal correlation with a coincidence-to-accidental ratio up to 16262$\pm$850. Our results confirm the feasibility of shallow etching in fabrication of efficient SPDC device on platform of lithium niobate on insulator, and benefit quantum information processing with broadband photon source.    
\end{abstract}

\maketitle

\section{Introduction}
Generating photon pairs via spontaneous parametric down-conversion~(SPDC)~\cite{Burnham1970PRL} is one of the most favorable techniques for quantum information processing with photons, which significantly facilitates quantum information processing with photons in the past three decades such as quantum cryptography~\cite{1993_Bennett_PhysRevLett.70.1895,2019_flamini_Rep.Prog.Phys._Photonica} and quantum computation~\cite{1998_steane_Rep.Prog.Phys._Quantum,2007_obrien_Science_Opticala}. The second-order optical nonlinearity lies at the heart of SPDC. Lithium niobate~(LN) is a promising material of choice for photon-pair generation owing to its strong optical susceptibility~($\chi^{(2)}$) and large transparency window~(from the ultraviolet to the mid-infrared)~\cite{1985_weis_Appl.Phys.A_Lithium}. More importantly, LN's ferroelectric property allows domain engineering (poling) and thus enables quasi-phase matching~(QPM), which is a crucial step for utilizing the second-order optical nonlinearity of LN. The thin-film LN on insulator~(hereafter referred as LNOI) platform---along with nano-fabrication technologies and periodical poling---significantly enhances the efficient generation of photon pairs for integrated quantum optics~\cite{Saravi2021AOM,2021_zhu_Adv.Opt.Photon._Integrated,Milad2022AP,Pelucchi2022NRP}, i.e., the QPM with the strong mode confinement in LNOI waveguide enables efficient SPDC with lower pump power compared to spontaneous four-wave mixing based on third-order nonlinearity~\cite{2021_wang_AppliedPhysicsReviews_Integrated,Milad2022AP}. Along this line of research, tremendous efforts have been devoted to demonstrate high-quality photon-pair generation on  LNOI platform with different structures, including micro-disk resonator~\cite{Frank2016,Luo2017OE}, ring resonator~\cite{Ma2020PRL} and periodically poled lithium niobate~(PPLN) straight waveguide~\cite{2019_chen_OSAContinuum_Efficient,2019_elkus_Opt.Express_Generation,Elkus2020OE,2020_zhao_Phys.Rev.Lett._Higha,Xue2021PRApplied,javid2021_Phys.Rev.Lett._Ultrabroadband, Xin22OL,2023_henry_Opt.Express_Correlated,2023_zhang_Optica_Scalable}.

The engineering of group velocity dispersion~(GVD) plays an important role in broadband nonlinear optical process, including SPDC~\cite{Xue2021PRApplied,javid2021_Phys.Rev.Lett._Ultrabroadband}, sum-frequency generation~\cite{Li17OL}, second harmonic generation~(SHG) and supercontinuum
generation~\cite{Jankowski20optica}. In particular, broadband photon pair from SPDC is promising for emerged quantum technologies, such as quantum imaging~ \cite{lemos2014_Nature_Quantum,kalashnikov2016_NaturePhotonics_Infrared,paterova2018_NewJ.Phys._Measurement}, frequency multiplexing~\cite{2017_grimaupuigibert_Phys.Rev.Lett._Heralded,2018_joshi_NatCommun_Frequency}, high-dimensional encoding~\cite{2010_olislager_Phys.Rev.A_Frequencybin,2010_sheridan_Phys.Rev.A_Security} and quantum optical coherence tomography~\cite{2002_abouraddy_Phys.Rev.A_Quantumoptical,2022_hayama_Opt.Lett._Highdepthresolution}. Moreover, the dispersion engineering of PPLN waveguide is able to generate photon pairs with high purity~\cite{Xin22OL}, which is promising for multiplexing single photon sources~\cite{Kaneda2019SA}. In LNOI-based SPDC, deep-etched waveguide is generally preferred due to the tighter mode confinement, consequently leading to higher conversion efficiency. However, deep-etched LNOI waveguide requires the low-roughness etching of the sidewall, which has only been achieved very recently, and is not yet broadly available~\cite{2021_zhu_Adv.Opt.Photon._Integrated}. In contrast, shallow-etched LNOI waveguide can reduce the scattering loss induced by the sidewall roughness. Moreover, shallow etching increases the tolerance for fabrication-induced discrepancy of waveguide geometry and relaxes the constraints on the required poling period, which has been demonstrated for highly-efficient SHG~\cite{2020_zhao_Opt.Express_Shallowetched}. 

In this work, we design and fabricate a 5.7-mm-long PPLN waveguide on a $600~\text{nm}$-thick $x$-cut LNOI with shallow etching of $165~\text{nm}$. The PPLN waveguide is able to efficiently generate photon pairs with wide bandwidth and strong temporal correlation. We observe photon-pair generation with brightness of 11.7~GHz/mW, bandwidth of 22~THz and coincidence-to-accidental ratio~(CAR) up to 16262.  

\begin{figure}[htb]
\centering
\includegraphics[width=\linewidth]{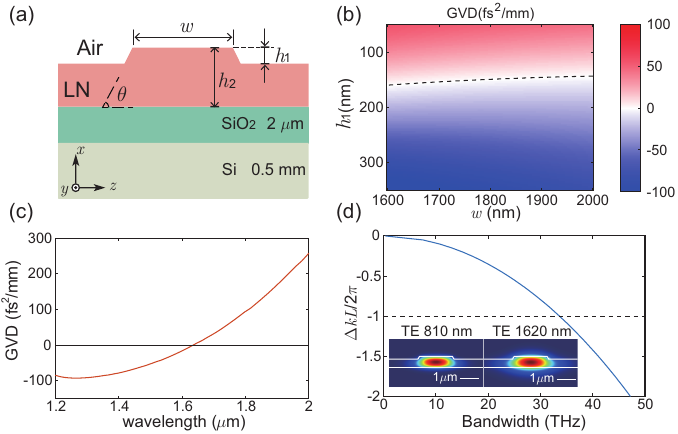}
\caption{Numerical simulation. (a) Cross section of the ridge waveguide on $h_2=600$~nm thick $x$-cut LNOI wafer, with the top width of $w=1800~nm$, a sidewall angle of $\theta=60^{\circ}$, an etch depth of $h_1=165~nm$. (b)Simulated the group velocity dispersion~(GVD) $k^{\prime\prime}$ with changes of  $w$ and $h_1$ in the cross section of ridge waveguide, where the wavelength is fixed at 1620 nm. The dashed black line denotes $k^{\prime\prime}=0$. (c) Simulated GVD with changes of wavelength where the cross section of ridge waveguide is fixed by $w=1800~nm$ and $h_1=165~nm$. (d) The phase mismatching $\Delta kL/2\pi$. Dashed line indicates the first zero of the phase-mismatch sinc function. The insets in (d) show the fundamental TE mode profiles of the waveguide at $810~\text{nm}$ and $1620~\text{nm}$.} 
\label{Fig:design}
\end{figure} 

\begin{figure}[ht!b]
\centering
\includegraphics[width=\linewidth]{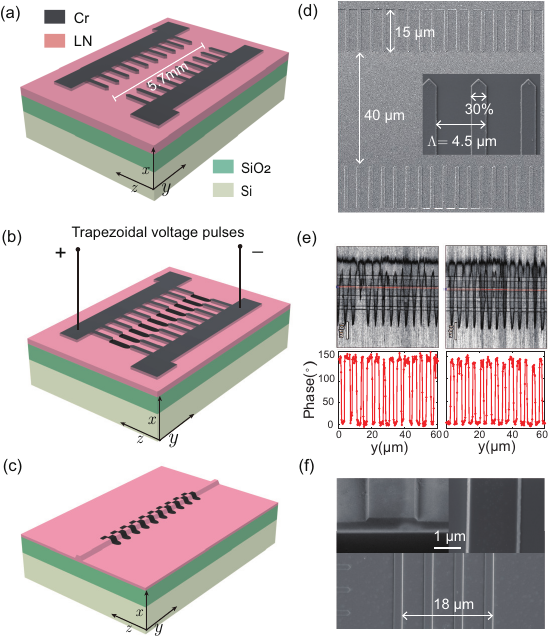} 
\caption{The fabrication of PPLN waveguide consists of three steps: (a) Pattern electrodes on LNOI surface, (b) Apply high-voltage pulses for periodically poling and (c) Etch the LN layer to form ridge waveguide. (d) The SEM images of the electrodes. (e) The PFM images of poling region. (f) The SEM images of the fabricated waveguides.} 
\label{Fig:fab}
\end{figure}

\section{Design and fabrication}
In SPDC, a short-wavelength pump photon~($p$) is spontaneously split into a pair of signal~($s$) and idler~($i$) photons of longer wavelengths. The three-photon mixing has to conserve energy $\omega_p=\omega_s+\omega_i$ with $\omega$ being the frequency. More importantly, phase-matching condition $\Delta k=k_s+ k_i - k_p=0$ of the three interacting modes must be met for efficient nonlinear interaction, where $k=\frac{2\pi}{\lambda}\cdot n_\text{eff}$ is the wavevector. To realize broadband SPDC, near-zero~(GVD$\approx$0) dispersion at the center of SPDC spectrum is required, which can be achieved by careful design of the geometry of waveguides. The LNOI sample we considered consists of 600-nm-thick $x$-cut LN thin film bonded to 0.5-mm-think Si substrate with 2-$\mu$m-think SiO$_2$. The cross-section of a typical ridge waveguide is shown in Figure~\ref{Fig:design}~(a),  which enables the type-0 phase-matching SPDC~($\text{TE}\to \text{TE}+\text{TE}$) process from pump light of 810~nm to degenerated signal and idler photon of 1620~nm with transverse-electric~(TE) modes in the waveguide. The GVD at frequency of $\omega_{s(i)}$ is $k^{\prime\prime}=\left(\frac{\partial^2 k}{\partial \omega^2}\right)_{\omega=\omega_{s(i)}}$, and can be numerically obtained by simulating the effective index $(n_\text{eff})_{s(i)}$ of TE modes of a waveguide with fixed width $w$ and etch depth $h_1$. As shown in Figure~\ref{Fig:design}~(b), near-zero GVD is achieved with shallow-etched depth of 165~nm and width range from 1600~nm to 2000~nm, which can tolerate the fabrication-induced variance of width $w$ in practice. With the simulated $n_\text{eff}$ by setting $w=1800~\text{nm}$ and $h_1=165$~nm, we calculate GVD of transmitted light with wavelength $\lambda_{s(i)}$ from 1.2~$\mu$m to 2~$\mu$m in waveguide. As shown in Figure~\ref{Fig:design}~(c), we observe the GVD $k^{\prime\prime}\leq100~\text{fs}^2/\text{mm}$ from $\lambda_{s(i)}=1.2~\mu$m to $\lambda_{s(i)}=1.8~\mu$m, which implies the capability of high conversion efficiency within this wavelength range. The dispersion-induced phase mismatching of $\Delta k$ is compensated by periodical poling of LNOI, where the orientation of the ferroelectric domains can be inverted by applying high voltage pulses in a process called poling~\cite{2016_chang_Optica_Thin,2016_mackwitz_Appl.Phys.Lett._Periodic,2018_nagy_Conf.LasersElectro-Opt._Periodic,2020_nagy_Opt.Mater.Express_Submicrometer,2020_zhao_J.Appl.Phys._Poling}. The poling period $\Lambda$ is calculated by 
\begin{equation}
\Delta k=\frac{2\pi}{\lambda_p}(n_\text{eff})_p-\frac{2\pi}{\lambda_s}(n_\text{eff})_s-\frac{2\pi}{\lambda_i}(n_\text{eff})_i-\frac{2m\pi}{\Lambda}=0,
\end{equation} 
In our SPDC device, the poling period $\Lambda\approx 4.5~\mu$m satisfies the first-order~($m=1$) QPM in type-0 SPDC of 810~nm$\to1620~\text{nm}+1620~\text{nm}$. By setting $\Lambda=4.5~\mu$m and $L=5.7$~mm, we calculate the phase mismatching $\Delta kL/2\pi$ as shown in Figure~\ref{Fig:design}~(d) which indicates the largest phase-matching bandwidth of 34~THz. We also calculate the mode profiles of the fundamental TE mode at 810~nm and 1620~nm in the waveguide with geometry shown in Figure~\ref{Fig:design}~(a), and the results are shown in inserts of Figure~\ref{Fig:design}~(d). The corresponding modal effective areas are $A_\text{eff}=0.8\mu \text{m}^2$ at 810~nm and $A_\text{eff}=1.3\mu \text{m}^2$ at 1620~nm, respectively. The normalized mode overlap is 93\%, which indicates the high-efficiency generation of the photon pairs. Indeed, the shallow etching may lead lateral mode leakage at the short wavelength~\cite{boes2019_Opt.Express_Improved}. We numerically calculate the transition loss 0.003~dB/cm of TE mode at 810~nm, which indicates the mode leakage of the pump light can be neglected. Also, according to the criterion proposed in Ref.~\cite{boes2019_Opt.Express_Improved}, we calculate the effective indexes of the TE mode in the waveguide and the TM mode in the slab and obtain the difference $\Delta n_\text{eff}=2.099-2.091>0$, which further confirms there is no mode leakage of pump light.

To fabricate the designed PPLN waveguide, we first periodically pole a 600-nm-thick $x$-cut LNOI wafer~(NANOLN Inc.). As shown in~Figure~\ref{Fig:fab}~(a), two comb-like chromium~(Cr) electrodes, with thickness of 100~nm and length of 5.7~mm along $y$ direction, is patterned on the surface of LNOI by deposition and lift-off process via electron beam lithography~(EBL). The gap width of two opposite electrodes is 40~$\mu$m. The triangle-tipped tooth of comb are fabricated with length of $15~\mu$m and period of  $4.5~\mu$m with $30\%$ duty cycle. Note that there is unavoidable lateral dispersion (along $y$ direction) during the poling process. The design of $30\%$ duty cycle and large gap width 40~$\mu$m provides flexibility to select poling region close to $50\%$ duty cycle, which corresponds to the maximal converting efficiency. The domain inversion between each pair of opposite teeth is realized by applying high-voltage pulses to the electrodes with two electrical probes as shown in Figure~\ref{Fig:fab}~(b). During the poling process, the poling region is immersed in silicone oil to avoid air breakdown. A multi-pulse voltage trapezoidal waveform with short pulse duration is applied on the poling pads, which is similar to Ref.~\cite{2016_chang_Optica_Thin,2018_wang_Optica_Ultrahighefficiency,2019_zhao_Opt.Express_Optical}. As shown in~Figure~\ref{Fig:fab}~(f), the domain inversion in two regions consisting of 24 periods are characterized by piezoresponse force microscopy~(PFM), where the dark grey regions correspond to the inverted domains. The inverted domains spread quasilaterally from the positive electrodes towards the negative electrodes, so that the location of waveguide should be carefully defined for high-efficiency SPDC. After poling, the electrodes are removed by metal etchant. As shown in Figure~\ref{Fig:fab}~(c), a second EBL is used to create waveguide patterns inside the poled regions, which is further transferred to the LN layer using reactive ion etching processing to form the ridge waveguide. We fabricate four ridge waveguides following this processing, and we find that the best poling profile is achieved on the second waveguide~(highlighted with a red line in~Figure~\ref{Fig:fab}~(f)). According to the results of PFM on this waveguide, we calculate that the duty of inverted domain is $0.50\pm0.12$. Before testing the SPDC, we characterize the classical nonlinear conversion efficiency of fabricated PPLN waveguide by SHG, which is with peak conversion efficiency of about 975$\%/\text{W}/\text{cm}^2$ (See Appendix~\ref{App:A} for details). In the following, the results of SPDC are obtained with this waveguide.

\begin{figure}[htb]
\centering
\includegraphics[width=\linewidth]{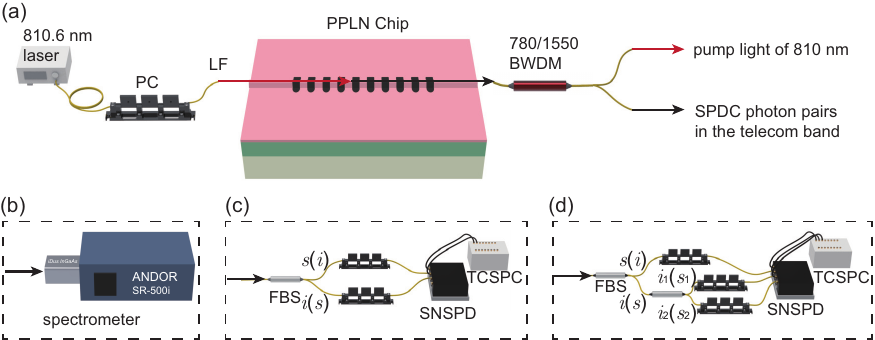} 
\caption{The experimental setups to characterize the generated photon pairs from PPLN waveguide. (a) The setup to pump PPLN waveguide. (b) The setup to measure the spectrum of signal photon. (c) The setup to measure the brightness and CARs of generated photon pairs. (d) The setup to measure the second-order autocorrelation function of heralded single photon. PC, polarization controller; LF, lensed fiber; 780/1550 BWDM, 780/1550~nm band wavelength division multiplexer; FBS, 50:50 fibered beamsplitter; SNSPD, superconducting nanowire single-photon detectors; TCSPC, time-correlated single-photon counting system.} 
\label{Fig:setup}
\end{figure} 

\section{Results of SPDC} 
As shown in~Figure~\ref{Fig:setup}~(a), a polarization controller~(PC) is used to control the polarization of a CW laser with central wavelength of 810.6~nm, which is further coupled into the PPLN chip by a lensed fiber to excite the TE mode of waveguide for type-0 SPDC. The PPLN chip is mounted on a thermoelectric cooler, and the temperature is set at 23$^\circ$ for phase matching. The generated signal and idler photons are coupled out from the PPLN chip with another lensed fiber, and the pump light is filtered out by a series of 780/1550~nm band wavelength division multiplexer~(BWDM). The end facets of PPLN chip are polished by chemical mechanical polishing~(CMP), and the fiber–chip–fiber loss is 35.4~dB at 810.6~nm and 18.1~dB at 1621.2~nm respectively.  

\begin{figure}[h!tbp]
 \centering
 	\includegraphics[width=\linewidth]{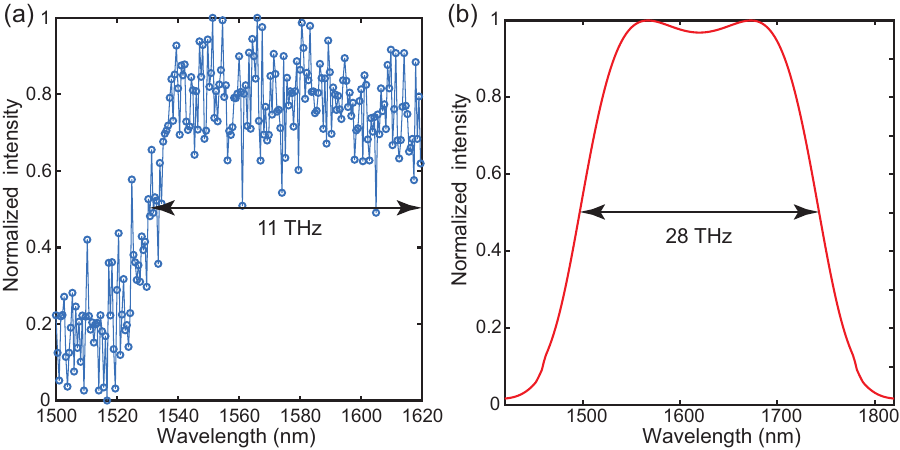} 
 	\caption{(a) The measured spectrum of signal photon. (b) The theoretical prediction of two-photon spectrum from joint spectrum intensity.} 
 	\label{Fig:spec}
\end{figure}

\begin{figure}[ht!bp]
\centering
\includegraphics[width=\linewidth]{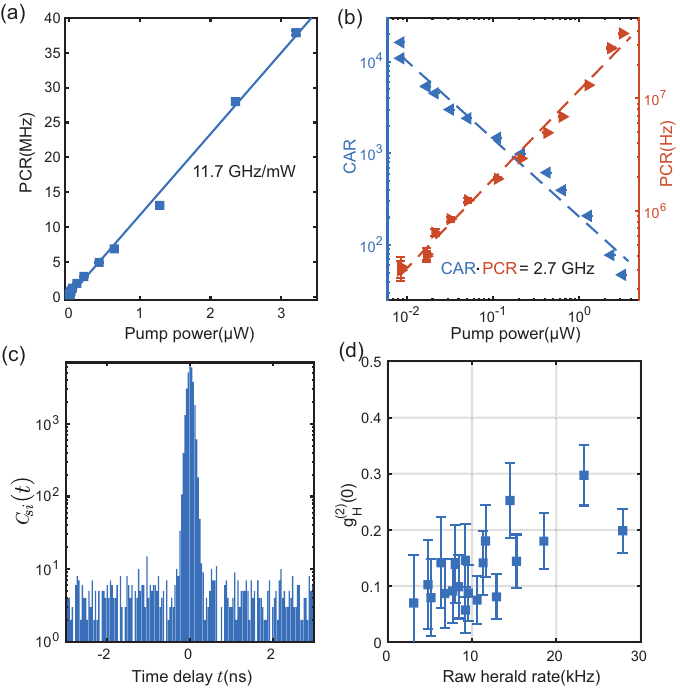} 
\caption{(a) PCRs with various pump power. (b) CAR and PCR with various pump power. The error bars are one standard deviation estimated by including the Poisson statistics of the emitted photons and the fluctuations. (c) The signal-idler coincidence $C_{si}(t)$ with different time delays $t$ with pump power of 0.11~$\mu$W. (d) The second-order autocorrelation function $g_{\text{H}}^{(2)}(0)$ of the heralded photon with various pump power. The error bars are the statistic errors obtained with Monte Carlo simulation by assuming the collected counts are with Poisson distribution.} 
\label{Fig:character}
\end{figure}

We first measure the bandwidth of generated photons, where the output lensed fiber is directly connected to an infrared spectrometer~(ANDOR, Shamrock SR-500i) equipped with  an InGaAs camera~(ANDOR iDus DU490A-1.7). The detection efficiency of the InGaAs camera is drastically decreased for the wavelength longer than 1.65~$\mu$m. Limited by this cutoff, we measure the spectrum shorter than 1621.2~nm, which is referred as signal photon. Then, the full bandwidth can be estimated according to energy conservation. As shown in Figure~\ref{Fig:spec}~(a), we obtain a spectrum accumulated in 10 minutes of signal photon with 3-dB half-bandwidth of 11~THz~(90~nm). According to the energy conservation $\omega_p=\omega_s+\omega_i$ in SPDC, the total bandwidth of signal and idler photons is 22~THz~(190~nm). Note that the bandwidth are highly related to the bandwidth of pump light. In our experiment, the bandwidth of pump light is 1.1~nm. Using the the bandwidth of pump laser and phase-matching condition $\Delta k$ in Figure~\ref{Fig:design}~(d), we calculate the joint spectrum intensity~(JSI) of SPDC which gives 28~THz bandwidth of the two-photon spectrum as shown in Figure~\ref{Fig:spec}~(b). The slight difference between experimental results and theoretical prediction is mainly attributed to the quality of periodical poling, which decreases the conversion efficiency for the signal~(idler) photon far from the center wavelength. 
We then test the brightness of photon-pair generation from PPLN waveguide using the setup illustrated shown in~Figure~\ref{Fig:setup}(c). The signal and idler photons are separated by a 50:50 fibered beam splitter~(FBS), detected by the superconducting nanowire single-photon detectors~(SNSPDs) located in a cryostat~(Quantum Opus, Opus One), and recorded by a time-correlated single-photon counting~(TCSPC) system~(Swabian Instruments, Time Tagger 20). We denote the counter rate of signal and idler photons as $C_s$ and $C_i$ respectively, and the coincidence as $C_{si}$. Then, the pair coincidence rate (PCR) is calculated by $C_iC_s/2C_{si}$, where the factor of 2 in the denominator is because of the FBS in the experimental configuration for detection. We measure the PCR with different values of input pump power, and the results are shown in~Figure~\ref{Fig:character}~(a). By linear fitting of the PCRs, we obtain a slope of 11.7~GHz/mW. Dividing by the bandwidth of 22~THz, the brightness of photon pair is $5\times10^5$~pairs/s/mW/GHz. We also measure the temporal correlation of the generated photons, which is determined by the coincidence-to-accidental ratio~(CAR), which is obtained by measuring the signal-idler coincidence with different time delay as shown in~Figure~\ref{Fig:character}~(b). The CAR is calculated by $\text{CAR}=\max[g_{si}^{(2)}(t)]-1$ with $g_{si}^{(2)}(t)=C_{si}(t)/C_{si}(\infty)$, where $C_{si}(t)$ is the coincidence with time delay $t$ and $C_{si}(\infty)$ is the coincidence with time delay far from $t=0$. We calculate the CARs with different value of pump power, and the results are shown with blue triangles in~Figure~\ref{Fig:character}(c). In contrast to the PCR, the CAR is inversely proportional to the power of pump light. In our experiment, the highest CAR of $16262 \pm 850$ is achieved with pump power of 8.6~nW, and the corresponding PCR is $296 \pm 60$~kHz. The product of CAR and PCR is independent of the pump power, and we obtain  $\text{CAR}\cdot \text{PCR}=2.7$~GHz according to~Figure~\ref{Fig:character}(c).

Finally, we characterize the property of heralded single-photon source~(HSPS) based on the SPDC, in which the detection of one photon is used to herald the emission of the other one. As shown in ~Figure~\ref{Fig:setup}~(d), the signal photon is detected to herald the emission of idler photon, and the idler photon is split by a 50:50 FBS. The nonclassical antibunching of HSPS can be characterized by second-order autocorrelation function of idler photon with zero time delay, i.e., $g_{H}^{(2)}(0)=C_{si_1i_2}C_{s}/2C_{si_1}C_{si_2}$, where $C_{si_1i_2}$ is the three-folder coincidence between $s$, $i_1$ and $i_2$. The results of $g_{H}^{(2)}(0)$ are shown in~Figure~\ref{Fig:character}~(d). At low pump power of 1.5~$\mu$W, we observe the HSPS with raw herald rate of 9.2~kHz and $g_{H}^{(2)}(0)=0.057\pm0.039$. By increasing the pump power, the herald rate increases while $g_{H}^{(2)}(0)$ decreases. At the highest power of 3.7~$\mu$W in this experiment, we observe HSPS with raw herald rate of 27.9~kHz and $g_{H}^{(2)}(0)=0.198\pm0.04$, which is still well below the classical threshold of 0.5.

\section{Conclusion}
In conclusion, we demonstrate the generation of photon pair with wide bandwidth, high brightness and strong temporal correlation, in a shallow-etched PPLN nanowaveguide. To the best of our knowledge, this is the first demonstration of SPDC with wide bandwidth covering telecom C-, L- and U-bands. A comparison of recent works focusing on generation of photon pair with wide bandwidth in PPLN straight waveguide are shown in Appendix~\ref{App:B}. Compared with previous shallow-etched PPLN waveguide~\cite{2019_elkus_Opt.Express_Generation}, our device shows the significant improvements in terms of brightness and CAR, which is comparable to the device with deep etching. The bandwidth of on-chip PPLN photon source can be further improved with chirped quasi-phase matching~\cite{Sensarn2010PRL}, which requires multiple poling periods.  Our results confirm the feasibility of shallow etching in fabrication of efficient SPDC device based on LNOI platform, and benefit quantum information processing with broadband photon sources.

\section*{Disclosures}
 The authors declare no conflicts of interest.

\section*{Data availability}
 Data underlying the results presented in this paper are not publicly available at this time but may be obtained from the authors upon reasonable request.

\bibliography{LNSPDC}

\appendix
\section{Results of SHG}
\label{App:A}
\begin{figure}[htb]
\centering
\includegraphics[width=\linewidth]{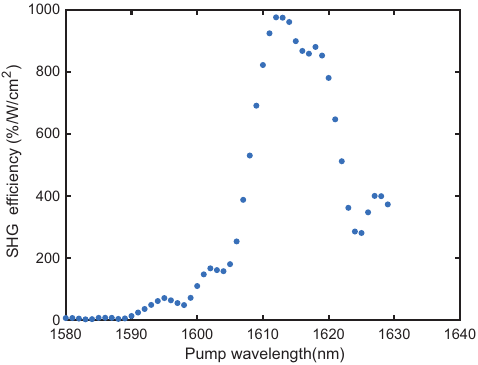} 
\caption{Experimental results of second harmonic generation of the fabricated PPLN waveguide.}
\label{Fig:SHG}
\end{figure}

We characterized efficiency of SHG $\eta$ by pumping the PPLN waveguide with a tunable laser at frequency of $\omega$ with power of $P_{\omega}$. By measuring the output power $P_{2\omega}$ at $2\omega$, we calculate the SHG efficiency $\eta$ by $\eta=\frac{P_{2\omega}}{L^{2} P_{\omega}^{2}}$, where $L=5.7$~mm is the length of PPLN waveguide. The experimental results are shown in Figure~\ref{Fig:SHG}, and the peak of the SHG efficiency we obtained is about 975\%/W/cm$^2$.

We also calculate the theoretical normalized SHG efficiency of the designed PPLN waveguide
~\cite{2018_luo_Optica_Highly}
\begin{align}
\eta=\frac{8 d_{33}^2}{\varepsilon_0 c n_\omega^2 n_{2 \omega} \lambda_{2 \omega}^2} \frac{\zeta^2}{A_{\mathrm{eff}}}\left[\frac{\sin (\Delta k\cdot L / 2)}{\Delta k\cdot L / 2}\right]^{2},
\end{align}
where nonlinear coefficient $d_{33}$=27~pm/V, $\varepsilon_0$ is the vacuum permittivity and $c$ is the speed of light in vacuum. By setting $\lambda_{\omega}$ = 1620~nm and $\lambda_{2\omega}$ = 810~nm, we calculate the effective refractive indices $n_\omega=1.92$ and $n_{2\omega}=2.099$, respectively. The effective model area is $A_{\text {eff }}=\left(A_{\omega}^2 A_{2{\omega}}\right)^{\frac{1}{3}}=1.106~\mu m^2$ where $A_{{\omega}}$ and $A_{2{\omega}}$ are the effective model profiles for TE mode at 1620~nm and 810~nm, respectively. $\zeta$ represents the spatial mode overlap factor between the two modes. $\Delta k=\frac{2\pi}{\lambda_{2\omega}} \left( n_{\omega}-n_{2\omega} \right)-\frac{2m\pi}{\Lambda}$ is the phase mismatching. Note that $\lim_{\Delta_k\to0}\left[\frac{\sin (\Delta k\cdot L / 2)}{\Delta k\cdot L / 2}\right]=1$. In the case of phase matching~($\Delta_k=0$), we calculate $\eta=$3364\%/W/cm$^2$.

\section{Comparison of broadband SPDC with PPLN}
\label{App:B}
In this section, we compare recent works aiming to generate broadband SPDC with PPLN straight waveguide. In this comparison, we focus the on-chip generation with PPLN straight waveguide. Other technique, such as chirped quasi-phase matching in bulk PPLN crystal~\cite{Sensarn2010PRL}, is beyond the scope of this comparison. 
\begin{table*}[htbp]
\renewcommand\arraystretch{2}
\centering
\begin{threeparttable}
\caption{Comparison of recent works focusing on generation of broadband photon pairs in PPLN waveguides, in terms of etching depth, bandwidth, brightness and CAR.}\label{Tb:comparasion}
    \begin{tabular}{c|c|c|c|c}
    \hline\hline
      & Etching Depth & Bandwidth & Brightness &CAR \\ 
     \hline
    This work
    & 165~nm 
    & 22~THz
    & 11.7~GHz/mW
    & 16262 \\
    \hline
    Elkus~\emph{et al.}~\cite{2019_elkus_Opt.Express_Generation}
    & 100~nm
    & 17.6~THz\tnote{b}
    & 50~MHz/mW\tnote{e}
    & 6900 \\
    \hline
    Xue~\emph{et al.}~\cite{Xue2021PRApplied}
    & 350~nm
    & 22~THz\tnote{c}
    & 279~GHz/mW
    & 599\\
    \hline
    Javid~\emph{et al.}~\cite{javid2021_Phys.Rev.Lett._Ultrabroadband}
    & 300~nm
    & 100~THz
    & 7.8~GHz/mW\tnote{f}
    & 20248\\
    \hline
    Henry~\emph{et al.}~\cite{2023_henry_Opt.Express_Correlated}
    & 350~nm\tnote{a}
    & 21~THz
    & 4.8~GHz/mW\tnote{g}
    & 7680\\
    \hline
    Zhang~\emph{et al.}~\cite{2023_zhang_Optica_Scalable}
    & 2.5~$\mu$m
    & 24~THz\tnote{d}
    & 178~MHz/mW
    & 8136\\
    \hline
    \hline
    \end{tabular} 
    \begin{tablenotes}
    \footnotesize
    \item[a] Hybrid SiN-LNOI waveguide, where the SiN is etched by 350~nm.
    \item[b] Calculated from spectrum with central wavelength of 1544.4~nm and bandwidth of 140~nm. 
    \item[c] Calculated from spectrum with central wavelength of 1475~nm and bandwidth of 160~nm. 
    \item[d] Calculated from spectrum with central wavelength of 1578~nm and bandwidth of 200~nm. 
    \item[e] Calculated from normalized brightness of $398 \times 10^6$~pairs/(s·nm·cm$^2$·mW), length of poled region $L=300~\mu$m and bandwidth of 140~nm. 
    \item[f] Calculated from Fig. 4~(c) in Ref.~\cite{javid2021_Phys.Rev.Lett._Ultrabroadband}.
    \item[g] Calculated from the PCR of 23~MHz/mW with 100~GHz filtering and the bandwidth of 21~THz.
    \end{tablenotes}
    \end{threeparttable}
\end{table*}

\end{document}